\documentstyle[preprint,aps,eqsecnum]{revtex} 
\input{epsf}
\def\be{\begin{equation}} 
\def\ee{\end{equation}}
\def\bea{\begin{eqnarray}} 
\def\eea{\end{eqnarray}}
\def\line{\hbox to \hsize}    
\def\frac #1#2{{#1\over #2}}

\def\det{{\rm det\,}}

\def \v{{\bf v}}

\def\vev #1{{\langle #1\rangle}}
\def\1{\mbox{\bf 1}}

\begin{document}
\draft 

\title{
Phonons and Forces: Momentum {\it versus\/} Pseudomomentum in Moving
Fluids
}

\author{ MICHAEL STONE}
\address{University of Illinois, Department of Physics\\ 1110 W. Green St.\\
Urbana, IL 61801 USA\\E-mail: m-stone5@uiuc.edu}   

\maketitle

\begin{abstract}

I  provide a pedagogical introduction to the notion of   pseudomomentum
for  waves in a medium, and show how changes in pseudomomentum may sometimes 
be used   to compute real forces. I then explain  how these ideas
apply to sound waves in a fluid. When the background fluid is in motion,
the conservation laws for pseudomomentum and pseudoenergy are most
easily obtained by exploiting the acoustic metric and the formalism of general
relativity.

\end{abstract}


\section{Introduction}

This conference  is devoted to  the physics of waves 
moving through a medium which affects them as would a 
background metric. There is therefore a natural analogy with
waves propagating in  a gravitational field --- but we
should take care not to push the analogy too far. These systems
differ from  real general relativity  in that the
medium constitutes a physical {\ae}ther. While we
may  ignore the {\ae}ther for many purposes, occasionally it is important.
For example,   if we wish to
compute forces exerted by the waves, we must  take
into account  any stress  transmitted by  the background medium.  

The natural tool for computing forces by  tracking the flux of 
energy and momentum  is the
energy-momentum tensor. This is best defined as the functional
derivative of the action
with respect to the background metric --- but
we   have two metrics at our disposal: the spacetime
metric and  the acoustic or other metric which we are
exploiting for our GR analogy. Despite the temptation to
believe otherwise, we should remember that it is only
by differentiating with respect to the ``real'' 
metric that we obtain  ``real'' energy and ``real'' momentum.
When we differentiate with respect to the analogy metric, 
we obtain  the density and flux of  other quantities. These  are usually   the
pseudoenergy and the pseudomomentum\cite{peierls79}. 

Failure to distinguish between real energy and momentum and
pseudoenergy and pseudomomentum has caused much confusion
and  controversy over the years. Consider the 
formula for the momentum of a photon in a dielectric:
should the refractive index go in the numerator, where it
was placed by  Minkowski, or in the denominator, as  argued
by Abrahams? This dispute was not resolved until
Blount\cite{blount71} identified    Minkowski's expression
with the pseudomomentum and   Abrahams'   with the true
momentum,  including the mechanical momentum of the
dielectric\cite{gordon73,nelson91}.  

Although I will not address the problem here, my initial
motivation for  thinking about these topics  was the fear
that a similar confusion  lies behind some recent
controversies\cite{stone00a} involving the Iordanskii
force. This  force, which has  an appealing GR analogue in
the gravitational Aharonov-Bohm effect of a spinning cosmic
string\cite{volovik98}, is supposed to act  on a vortex in
a superfluid when it moves relative to the normal component
of the fluid. 

A related issue, and one  that I will address, lies at the heart of 
the two-fluid model for a superfluid or Bose condensate. It
is one of the fundamental assumptions of the two-fluid
model that the  phonons in a fluid possess  momentum $\hbar
{\bf k}$, and that, unlike that of phonons in a solid, this
momentum is  true Newtonian momentum, $m{\bf v}$. This
assumption is essential because we wish to identify the phonon
momentum density with the mass current, both being equal
to ${\rho {\bf v}}$. The desired identification is supported by 
the approximate solution of Bogoliubov's weakly
interacting  Bose gas model, in which the  phonon creation
operator, $\hat a^\dagger_{\bf k}$, appears to be  create
bona-fide momentum, so it is quite unnerving to discover
that in the literature of fluid mechanics the attribution
of real momentum to a sound wave is regarded as a
na{\"\i}ve and dangerous fallacy. A particularly forceful
statement of this opinion is to be found in the
paper\cite{mcintyre81} {\it On the Wave Momentum Myth\/},
by Michael McIntyre. 

In the present  article we will focus   the difference between the
true momentum and the pseudomomentum associated with the
acoustic metric. The first part is a general
pedagogical account  of the distinction between momentum 
and pseudomomentum, and the circumstances under which the
latter may be used for computing forces. The second
part will discuss  the  energy and momentum associated with
sound waves in a background flow\cite{stone00b}.

\section{Momentum and Pseudomomentum} 

The distinction between true momentum and
pseudomomentum is especially clear when we  
consider the problem of transverse  vibrations  on an
elastic string. The action  
\be
S=\int dx\, dt\left\{\frac \rho2 \dot y^2 -\frac
T2  y'^2 \right\}
\ee
gives rise to the familiar wave equation
\be 
\rho \ddot y -Ty''=0,
\ee
with $c=\sqrt{T/\rho}$ being the wave speed.
By manipulating  the wave equation  we can establish two local conservation
laws. The first,
\be
\frac{\partial}{\partial t} \left\{ \frac \rho2 \dot y^2 +\frac
T2  y'^2\right\}+
\frac{\partial}{\partial x} \left\{-T\dot y y'\right\}=0,
\ee
is immediately recognizable  as an energy
conservation law, with the flux  $-T\dot y y'$
being the rate of doing work by an element of the
string on its neighbor. The second, 
\be
\frac{\partial}{\partial t} \left\{-\rho \dot y y'\right\}+
\frac{\partial}{\partial x} \left\{\frac \rho 2 \dot y^2 +\frac
T2 y'^2\right\}=0,
\ee
is slightly more obscure in  its interpretation. 

In a relativistic system the appearance of  $-\rho
\dot y y'=$ (energy flux)$/c^2$  as the local density of a
conserved quantity would not be surprising. The symmetry of
the energy-momentum tensor requires that  $T^{j 0}=T^{0
j}$,   so the   energy flux $T^{j 0}$ is also (after
division  by $c^2$) the density of 3-momentum. Here,
however, we are dealing with  non-relativistic classical
mechanics --- and with transverse waves.   Whatever the
quantity $-\rho \dot y y'$ may be, it is {\it
not\/} the density of the $x$ component of the  string's  momentum.  It is
instead the density of {\it pseudomomentum\/}.

To understand the origin of pseudomomentum, observe that our 
elastic string may be subjected to  two quite   distinct 
operations either of which  might be called ``translation in the $x$ direction'':

\begin{itemize}   

\item An operation  where the string, together with  any disturbance on
it, is  translated in the $x$ direction. 

\item An operation  where the string itself is left fixed,
but the disturbance is translated in the $x$ direction.

\end{itemize}

The first operation leaves the action invariant provided space
is homogeneous.  The associated conserved quantity is true 
Newtonian momentum.  The second operation  is a symmetry
only when both the  background space  {\it and the  string\/} 
are  homogeneous. The conserved quantity here is 
pseudomomentum.  Such a distinction between true and pseudo-
momentum is  necessary whenever  a medium
(or {\ae}ther) is involved. 

Adding to the confusion is that, although the pseudomomentum
is conceptually distinct from the the true momentum, there are
many circumstances in which changes in pseudomomentum can
be used to compute real forces.  As an example consider a
high speed train picking up its electrical power from an
overhead line.

\vbox{
\vskip 20 pt

\centerline{\epsfxsize=5.0in\epsffile{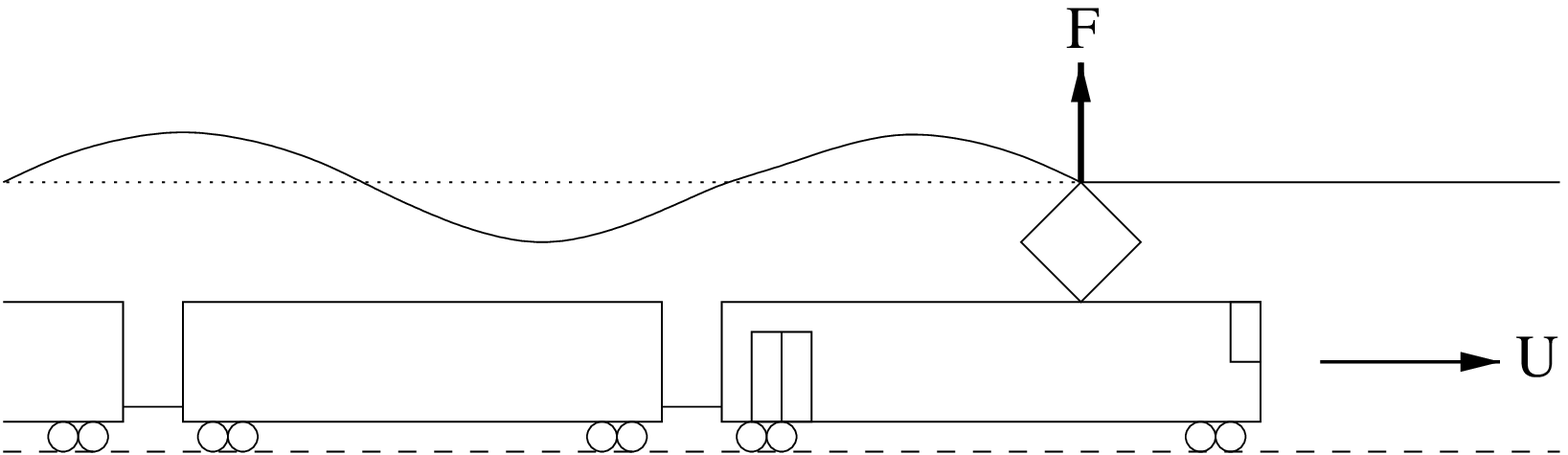}}
}
{\narrower\smallskip\centerline{\sl Fig 1. A high-speed train .
\smallskip}}

The locomotive is travelling at speed $U$ and the 
pantograph pickup is  exerting a constant vertical
force $F$ on the power line. We make the usual small
amplitude approximations and assume (not unrealistically) 
that the  line  is supported in such a way that its
vertical displacement obeys an inhomogeneous  Klein-Gordon
equation
\be
\rho \ddot  y -T y'' +\rho\Omega^2 y =
F\delta(x-Ut),
\label{EQ:locoKG}
\ee
with $c= \sqrt{T/\rho}$, the  velocity
of propagation of short wavelength disturbances.

If $U < c$, the
vertical displacement relaxes  symmetrically about the point
of contact.  Once  $U$ exceeds $c$, however, the character
of the problem changes from elliptic to hyperbolic, and an
oscillatory  ``wake'' forms behind the pantograph.  
As with all such wakes, the disturbance  is stationary when viewed
from   the frame of the train. With this in mind, we seek a
solution to (\ref{EQ:locoKG}) of the form $y=y(x-Ut)$. Since
the overhead line is undisturbed ahead of the locomotive, we find
\bea
y&=&\frac {\gamma F c}{\Omega T} \sin
\frac{\Omega\gamma}{c}(Ut-x),\qquad x<Ut \nonumber\\
y&=& 0,\qquad\qquad\qquad\qquad\qquad x>Ut.
\label{EQ:solution_y}
\eea
Here $\gamma= (U^2/c^2-1)^{-\frac 12}$ is the Lorentz
contraction factor
modified for tachyonic motion.  The condition that the
phase velocity, $\omega/k$, of the wave constituting the
wake be  equal to the  forward velocity of the object
creating it is analogous to  the Landau criterion
determining  the critical velocity of a superfluid. There
are no waves  satisfying this condition when $U<c$, but they 
exist for all $U>c$.      

In the wake, the time and space averaged energy density $\vev{\cal
E}=\vev{ \frac 12
\rho \dot y^2 +\frac 12 T y'^2 +\frac 12 \Omega^2 y^2}$ is
given by
\be
\vev{\cal E}= \frac 12 \rho \gamma^4 \left(\frac F
T\right)^2 U^2.
\ee
The expression for the pseudomomentum density for the
Klein-Gordon
equation is the same
as that for the wave equation, and the  average pseudomomentum density is
\be
\vev{-\rho \dot y y'}= \frac 12 \rho \gamma^4 \left(\frac F
T\right)^2 U.
\label{EQ:pmom_density}
\ee   

Because energy is being transfered from the locomotive to
the overhead line, it is clear that there must be some
induced drag, $F_{\rm d}$, on the locomotive. This is most
easily computed
from energy conservation. The rate of working by the locomotive, $F_{\rm
d} U$, must equal the energy density times the rate of
change of the length of the wave-train. Thus      
\be 
F_{\rm d}U=  \vev {\cal E}(U-U_g)
\ee
where 
\be
U_g= \frac {\partial\omega}{\partial k}= \frac
{c^2k}{\omega}= \frac{c^2}{U}
\ee
is the group velocity of the waves, and so 
the speed of the trailing end of the wake wave-train.

After a short calculation we find that the wave-induced drag force is 
\be
F_{\rm d}= \frac 12 \frac{F^2\gamma^2}{T}.
\ee 
Since the average pseudomomentum density turned out to be
the average energy density divided by $U$,  we  immediately
verify that we get exactly the same answer for $F_{\rm d}$
if we equate the wave drag to the time rate of change of
the total pseudomomentum.

For the sceptic we note that we  may also obtain the same
answer by a more direct evaluation of the force required to
deflect the  overhead wire.

\vbox{
\vskip 20 pt

\centerline{\epsfxsize=3.0in\epsffile{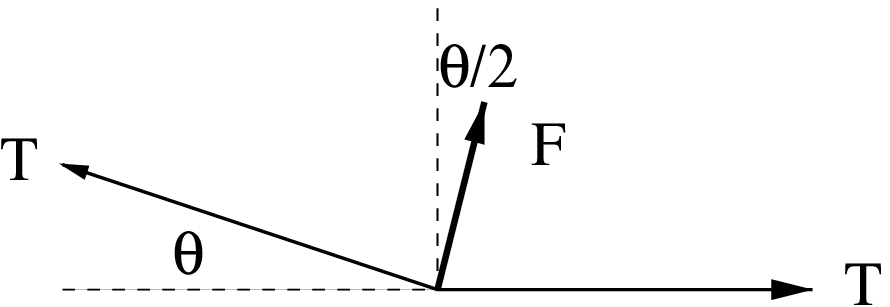}}
}
{\narrower\smallskip\centerline{\sl Fig 2. Force on overhead wire.
\smallskip}}

From the solution (\ref{EQ:solution_y}) we see that  the
force $F$ is related to the angle of upward
deflection, $\theta $, by $ \theta\approx \tan\theta =
-y'(Ut,t)=\gamma^2 F/T$. By balancing the acceleration of the power
line against the force and the tension in the  line, we see
that the  force  cannot be exactly vertical, but must be
symmetrically  disposed with respect to the horizontal and
deflected parts of the line. The force exerted by the pantograph 
thus has a
small horizontal component $F\sin \theta/2$. The wave drag is
therefore $F_{\rm d}= \frac 12 {F^2\gamma^2}/{T}$, as
found earlier.

Since there is a real horizontal  force acting on the wire, true
Newtonian momentum must also be being transfered to the
wire. Indeed the  wire behind the train is being stretched,
while that in front is being compressed. A section of
length $2c_{\rm long} (t-t_0)$, where $c_{\rm long}$ is the
velocity of {\it longitudinal\/} waves on the wire and
$(t-t_0)$ is the elapsed time, is in uniform motion in the $x$
direction.  Since usually $c_{\rm long}\gg c$, this true
momentum is accounted for by an almost infinitesimal motion
over a  large region of the wire. The pseudomomentum and
the true momentum are  to be found  in quite different
places --- but the divergence of their  flux tensors, and
hence the associated forces, are equal.

\section{Radiation Pressure}

In this section we will consider the ``radiation pressure''
exerted by sound waves  incident on an object immersed in
the medium. This is a subject with a long history of
controversy\cite{post53,beyer78}. The confusion  began  with the
great Lord Rayleigh who gave several inequivalent answers
to the problem.  Our discussion will follow that of Leon
Brillouin\cite{brillouin25} who greatly clarified the
matter. We begin by considering  some analogous situations
where the force is exerted by transverse waves on a string.

Consider a standing wave on a semi-infinite elastic string.

\vbox{
\vskip 20 pt

\centerline{\epsfxsize=3.0in\epsffile{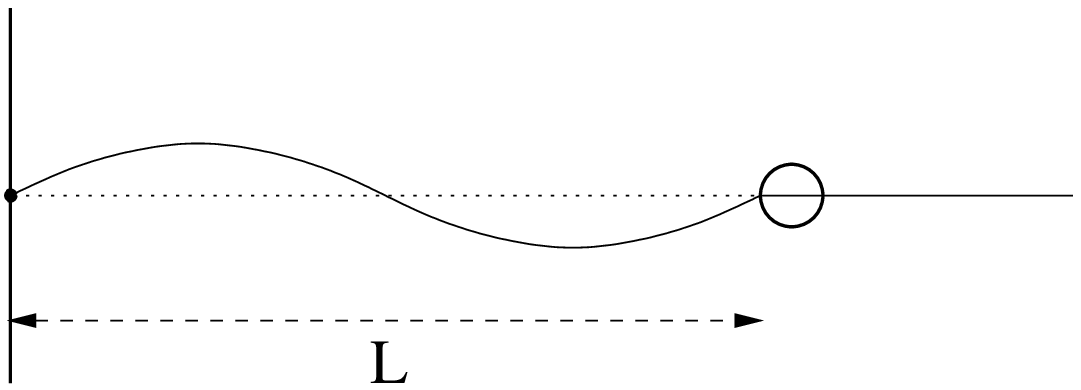}}
}
{\narrower\smallskip\centerline{\sl Fig 3. A vibrating string
exerts a force on a bead.
\smallskip}}

We have restricted the  vibration to the finite interval $[0,L]$
by means of a frictionless bead which forces the transverse
displacement of the string to be zero at $x=L$, but allows
free passage to longitudinal motion, and so does not affect
the tension. 

Suppose the  transverse displacement is 
\be
y=A \sin \omega t \sin \frac{\pi n x}{L}
\ee 
with $\omega = c\pi n/L$.
The total energy of the motion is
\be
E= \int dx\, \left\{ \frac 12 \rho \dot y^2 + \frac 12 T
y'^2\right\}= \frac 14 \rho\omega^2 A^2 L.
\ee  

If we alter the size of  the vibrating region by slowly moving the
bead, we will alter the energy in the oscillations. 
This change in energy  may be found by
exploiting the Boltzmann-Ehrenfest  principle which
says that  during an adiabatic variation of the parameters of 
a harmonic oscillator the quantity $E/\omega$ remains  constant. Thus
$ \delta(E/\omega)=0$ or $\delta E=
\left(\frac{E}{\omega}\right) \delta \omega$.
To apply this to the string, we note 
that
\be
\delta \omega = \delta \left( \frac{cn\pi}{L}\right) =-
\omega \frac {\delta L}{L}.
\ee
The change in energy, and hence the work we must do to
move the bead, is  then
\be
\delta E = - \left(\frac{E}{L}\right) dL.
\ee
The ``radiation pressure'' is therefore $E/L=\vev{\cal E}$,
the mean energy density. This calculation can be confirmed
by examining the forces on the bead along the lines of Fig.
2. 

 The average density of pseudomomentum in each of the
two  travelling wave components of the standing wave is  $
\pm\frac 12 \vev{\cal E}/c$. The   radiation ``pressure''
can therefore be accounted for by the $2\times c\times \frac 12
\vev{\cal E}/c$ rate of change of  the pseudomomentum in
the travelling waves as they bounce off the bead. Thinking
through this example shows why pseudomomentum can be used
to compute real forces: On a homogeneous string the act of
translating the bead and the wave together, while keeping
the string fixed,  leaves the action invariant. The
associated conserved quantity is the sum of the {\it
pseudomomentum\/} of the wave and the {\it true\/} momentum
of the bead.

Keeping track of pseudomomentum cannot account for all
forces, however.  There is another plausible way of
defining the radiation pressure. This time we use a finite string and
 attach its  
right-hand  end
to a movable wall

\vbox{
\vskip 20 pt

\centerline{\epsfxsize=3.0in\epsffile{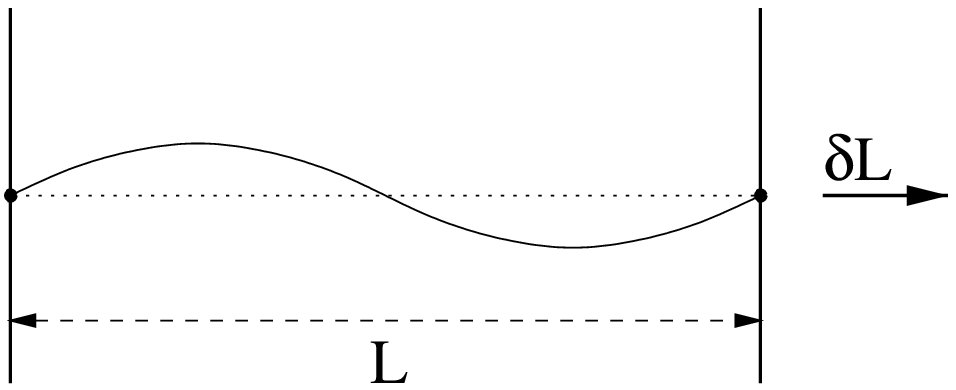}}
}
{\narrower\smallskip\centerline{\sl Fig 4. Another way to
define radiation pressure.
\smallskip}}

Now, as we alter the length of the string, we will change 
its tension, and so alter the value of
$c$. We must take the effect of this  into account in the 
variation of the frequency
\be
\delta \omega = \delta \left( \frac{cn\pi}{L}\right)=
-\omega \frac {\delta L}{L} + \omega \frac {\delta c}{c}.
\ee
The change in the energy of the vibrating system is
therefore 
\be
\delta E = -\vev{\cal E} \left(1- \frac {\partial \ln c}{\partial
\ln L}\right) \delta L.
\ee
The radiation pressure is thus  given by
\be
p= \vev{\cal E} \left(1- \frac {\partial \ln c}{\partial
\ln L}\right).
\ee
This force is in addition to the steady pull from the static
tension $T$ in the string.

The generality of the Boltzmann-Ehrenfest principle allows
us to apply  the previous discussion with virtually no changes to
compute forces exerted by a sound wave. We need no explicit
details of the wave motion beyond it being harmonic.  The
two ways of defining the radiation pressure for a vibrating
string correspond to two different experimental
conditions   that  we might use for measuring the 
radiation pressure for sound waves. The movable  end
condition corresponds to what is called the {\it
Rayleigh\/} sound pressure.     

\vbox{
\vskip 20 pt

\centerline{\epsfxsize=3.0in\epsffile{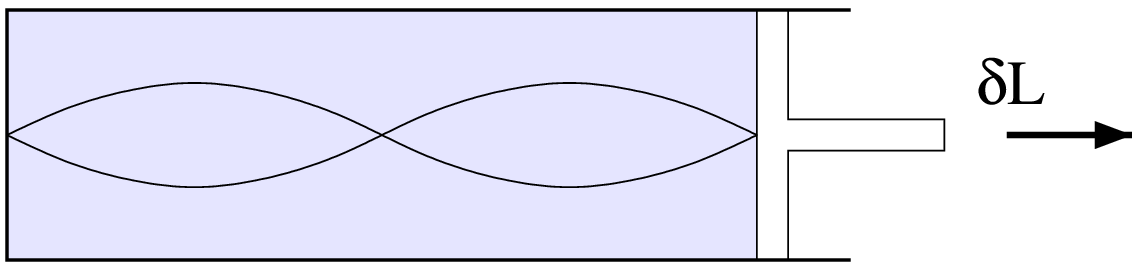}}
}
{\narrower\smallskip\centerline{\sl Fig 4. The Rayleigh Radiation Pressure.
\smallskip}}

We establish a standing wave in a  cylinder closed at one end and
having a movable piston at the other. 
Moving the piston confining the sound wave changes both the
wavelength of the sound and the speed of propagation, 
producing a sound radiation pressure
\be
p= \vev{{\cal E}}\left(1+ \left(\frac{\partial\ln  c}{\partial
\ln \rho}\right)_S\right).
\ee
The subscript $S$ on the derivative indicates that it is being 
taken at fixed entropy. As with the string,  this radiation pressure is in
addition to the equilibrium  hydrostatic pressure, $P_0$, on the piston.   

The analogue of the string with the sliding bead leads to
the {\it Langevin\/} definition of the radiation pressure.
Here we insert a  bypass so that moving the piston
confining the sound wave does not change the density or
pressure of the fluid. The radiation pressure on the
piston is simply $\vev{{\cal E}}$.

\vbox{
\vskip 20 pt

\centerline{\epsfxsize=3.0in\epsffile{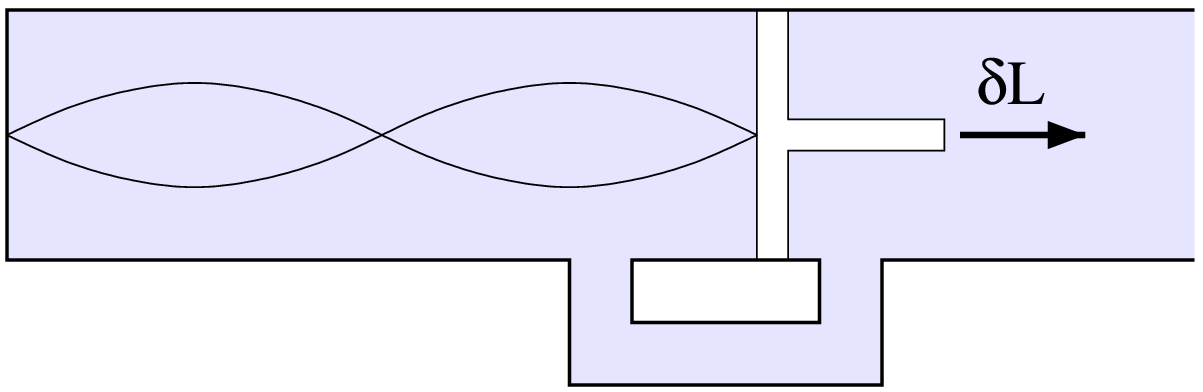}}
}
{\narrower\smallskip\centerline{\sl Fig 5. Langevin Radiation Pressure.
\smallskip}}

The difference in the two definitions  of radiation pressure arises because, 
if  we keep the mean pressure fixed, the  presence of a sound
wave produces  an
$O(A^2)$ change  in 
the  volume of the fluid. If, instead,  the mean density is held fixed, 
as it is in the Rayleigh definition, then this volume change 
is resisted by an  additional  hydrostatic pressure on the walls of 
the container.

The most common way of measuring sound pressure involves a sound
beam in an open tank of fluid. Since the fluid is free to expand, this 
corresponds to the Langevin pressure.   

\vbox{
\vskip 20 pt

\centerline{\epsfxsize=3.0in\epsffile{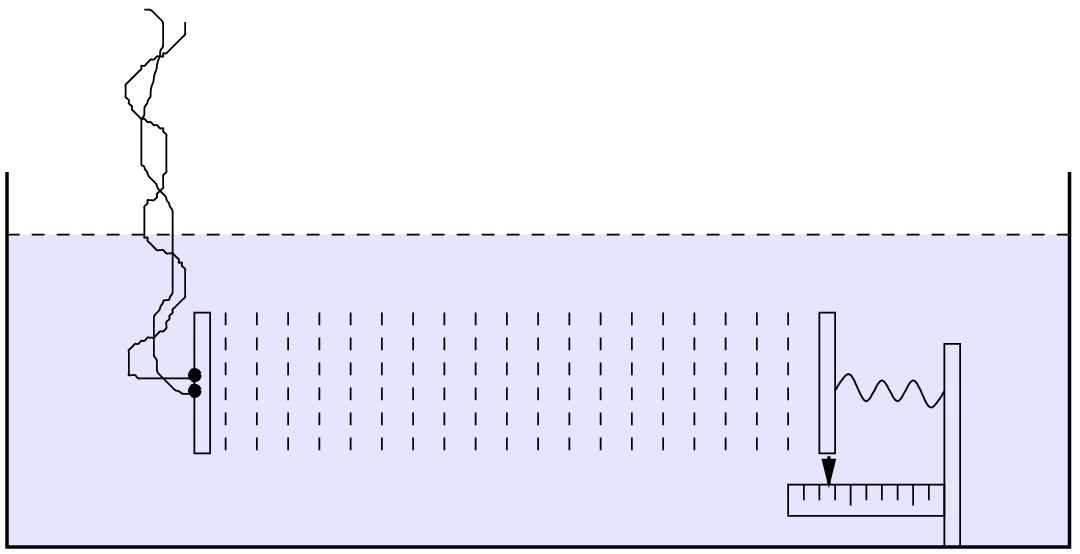}}
}
{\narrower\smallskip\centerline{\sl Fig 6. The usual experimental
situation.
\smallskip}}

The radiation pressure is, in reality, a radiation {\it
stress\/}
\be
\Sigma_{ij} = \vev{{\cal E}}\left( \frac{k_ik_j}{k^2}
+\delta_{ij} \left(\frac{\partial\ln  c}{\partial
\ln \rho}\right)_S\right).
\ee  
The anisotropic part  depends on the wave-vector ${\bf k}$ of the
sound beam, and may be  accounted for by keeping track of
the pseudomomentum changes. The isotropic part cannot be
computed from the linearized sound-wave equation since it
requires more information about the equation of state of the
fluid medium than is used in deriving the  wave
equation.   The extra information is encapsulated in the
parameter $\left(\frac{\partial\ln  c}{\partial
\ln \rho}\right)_S$, a fluid-state analogue of the
Gr{\"u}neisen parameter which characterizes the thermal
expansion of a solid. When the experimental situation is
such that this isotropic pressure is important, the force
associated with the sound field cannot be obtained from the
pseudomomentum alone.

\section{Mass Flow and the Stokes Drift}  

Further  confusion involving  momentum and pseudomomentum in
acoustics is  generated by  the need to  distinguish  between the Euler 
(velocity field at a particular  point) 
description of fluid flow, and the Lagrangian (following the  particles)
description.

Suppose the  velocity field in a sound wave  is 
\be
v_{(1)}(x,t) = A \cos (kx-\omega t).
\ee
Using the continuity equation 
$\partial_x \rho v+ \partial_t \rho=0$, setting
$\rho=\rho_0+\rho_{1}$, and approximating
$\rho v \approx\rho_0 v_{(1)}$, we find that
\be
\rho_1= \frac 1{\omega}\rho_0 k A \cos (kx-\omega t) + O(A^2).
\ee
The time average of the momentum density $\rho v$ 
is therefore
\be
\vev{\rho v} = \vev {\rho_1 v_{(1)}}= 
\frac {k\rho_0}{\omega} \frac 12 A^2, 
\label{EQ:mom_momav}
\ee
to $O(A^2)$ accuracy. This Newtonian momentum density is
clearly non-zero, and numerically equal to the
pseudomomentum density. Here, unlike the case of the elastic
string, there is only one velocity of wave propagation and so
the pseudomomentum and true momentum, although logically
distinct, are to be found in the same place.

Further  $\vev{\rho v}$ is both the momentum density 
and   the mass-current. A nonzero average
for the former therefore implies a 
steady drift of particles in the direction
of wave propagation, in addition to the back-and-forth
motion in the wave. We can confirm this by translating the Eulerian
velocity field  
$v_{(1)}= A \cos (kx-\omega t)$ into  Lagrangian language. 
The  trajectory
$\xi(t)$ of a particle intially at $x_0$ is the solution of the equation
\be
\frac {d \xi}{dt}= v_{(1)}(\xi(t),t)= A \cos (k\xi-\omega t),\qquad \xi(0)=x_0.
\label{EQ:mom_nle}
\ee
Since the quantity $\xi$ appears both in the derivative
and in the cosine, this
is a nonlinear equation. We solve it perturbatively by setting
\be
\xi(t) = x_0 + A\, \Xi_1(t) + A^2 \,\Xi_2(t) +\cdots.
\ee
We find that
\be
\Xi_1(t) = - \frac 1{\omega}\sin(kx_0-\omega t).
\ee
Substituting this into (\ref{EQ:mom_nle}) we find
\be
\frac {d\, \Xi_2}{dt}= \frac {k}{\omega}\sin^2(kx_0-\omega t).
\ee
Thus $\dot \Xi_2$ has a non-vanishing time average, $ k/(2\omega)$,
leading to a secular
drift velocity $\bar v_L =  \frac 12 kA^2/\omega $ that is consistent with (\ref{EQ:mom_momav}). This
motion is called the {\it Stokes drift\/}.

We also see why there is no net Newtonian momentum associated with
phonons in  a crystal. The
atomic displacements  in a harmonic crystal are given by
\be
\eta_n = A \cos(k(na)-\omega t),
\ee
so the   crystal equivalent of (\ref{EQ:mom_nle}) is
\be
\frac {d \eta_n}{dt}=A \cos (k(na)-\omega t).
\ee
Thus $\eta$ does not appear on the right-hand side of
this equation. It is  a  linear equation and gives
rise to  no net particle drift.   

So,  a sound wave does have real momentum? --- But wait!  The
momentum density we have computed is {\it second order\/} in the
amplitude $A$. The wave equation we have used to compute it is
accurate only to {\it first order\/} in $A$. 
We may  expand the velocity field as   
\be
\v= \v_{(1)}+\v_{(2)}+\cdots,
\ee
where the second-order correction $\v_{(2)}$ arises 
because the equations of fluid motion are non-linear. 
This correction will possess both oscillating and steady
components. The oscillatory  components arise  because a
strictly harmonic  wave with frequency $\omega_0$ will
gradually develop  higher frequency components due to  the
progressive distortion of the wave as it propagates. (A
plane wave eventually degenerates into a sequence of 
shocks.) These distortions  are usually not significant in
considerations of energy and momentum balance.   The steady
terms, however, represent $O(A^2)$ alterations to the  mean
flow caused  by the sound waves, and these may  possess 
energy and momentum  comparable to that of the  sound
field.

\vbox{
\vskip 20 pt

\centerline{\epsfxsize=3.0in\epsffile{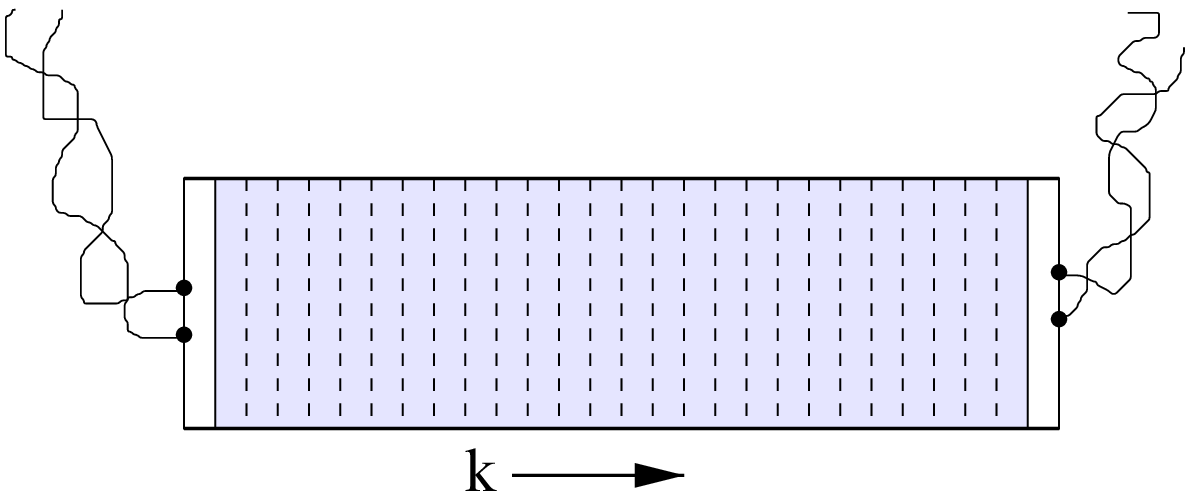}}

{\narrower\smallskip\centerline{\sl Fig 7. Momentum flux
without mass-flow.
\smallskip}}
}
For example, we   may drive a  transducer so as to produce
a beam of sound which totally fills a  closed cylinder of
fluid. At the far end of the cylinder  a  second transducer with suitably
adjusted amplitude and phase absorbs the beam without
reflection. Since the container is not going anywhere, it
is clear that the average velocity of the center of mass of
the fluid must be zero, despite the presence of the sound
wave. An exact solution of the non-linear equation of
motion for the fluid  must provide a steady component in 
$\v_{(2)}$    and this counterflow  completely cancels the
$\vev{\rho_1 v_{(1)}}$ term. Indeed in   Lagrangian
coordinates  the fluid particles simply oscillate back and
forth with no net drift. The wave momentum and its
cancelling counterflow are simply artifacts of our  Eulerian
description.  This is one reason  why professional fluid mechanics
dislike the notion of momentum being associated with sound
waves. 

The outlook for  the two-fluid model is not entirely bleak, however.     
The $\v_{(2)}$ corrections do not always exactly cancel the
momentum.  Any  non-zero value for $\nabla\cdot
\rho_1 \v_{(1)}$ --- such as occurs at the transducers at the ends of the   
cylinder ---  will act as a source or sink for a
$\v_{(2)}$ counterflow, but its exact form depends on the
shape of the container and other effects extrinsic to the
sound field. For transducers immersed in an infinite volume of fluid,
for example, the counterflow  will take the form of a
source-sink dipole field, and, although the total momentum of the
fluid will remain zero, there will be a non-zero momentum density.    
Because it is not directly associated with the sound field,  
in the
language of  the two-fluid  model  the  induced $\v_{(2)}$
counterflow is not counted as belonging to the  normal component of the
fluid, {\it i.e.\/} to the phonons,   but is lumped into
the background superflow. 
It is  determined by  enforcing mass conservation,  
\be 
\nabla
\cdot \rho_s {\bf v}_s+ \nabla \cdot \rho_n {\bf v}_n=0.   
\ee 
The difference of opinion between the physics and fluid mechanics 
communities over  whether phonons have real momentum
reduces, therefore, to one of different accounting
conventions.

\section{The Unruh wave equation}

Now we will examine  the energy and momentum in  a moving fluid. 
The flow of an irrotational fluid is derivable from the
action 
\cite{schakel96}
\be
S= \int d^4x\left\{ \rho \dot \phi +\frac 12 \rho (\nabla\phi)^2
+ u(\rho)\right\}.
\label{EQ:basic_action}
\ee
Here $\rho$ is the mass density, $\phi$ the velocity potential, and
the overdot denotes differentiation with respect to time. The
function $u$ may be identified with the internal energy density.  

Varying with respect to  $\phi$
yields the continuity equation 
\be \dot \rho + \nabla\cdot (\rho \v)=0,
\label{EQ:continuity} \
\ee
where $\v\equiv\nabla \phi$.
Varying
$\rho$ gives a form of Bernoulli's equation 
\be \dot\phi +\frac 12 \v^2 +
\mu(\rho)=0, 
\label{EQ:bernouilli} 
\ee where $\mu(\rho) = {d u}/{d \rho}$.
In  most applications $\mu$ would be identified with the specific
enthalpy.  For  a superfluid condensate  the entropy density, $s$, is
identically zero and  $\mu$ is the local chemical potential.

The gradient of the Bernoulli equation is 
Euler's equation of motion for the fluid. Combining this with the continuity equation
yields a  momentum conservation law
\be
\partial_t(\rho v_i)+\partial_j(\rho v_j v_i
) +\rho\partial_i \mu=0.
\label{EQ:momcons1} 
\ee
We simplify (\ref{EQ:momcons1}) by introducing the pressure, $P$, which is 
related to $\mu$ by   
$P(\rho)=\int\rho d\mu$. Then we can write 
\be
\partial_t(\rho v_i)+\partial_j \Pi_{ji}=0,
\ee
where   $\Pi_{ij}$ is
given by
\be 
\Pi_{ij}= \rho v_iv_j+\delta_{ij}P.
\ee
This is the usual form of the momentum flux tensor in  
fluid mechanics. 

The relations $\mu= {d u}/{d \rho}$ and $\rho ={d P}/{d \mu}$ show that  
$P$ and $u$ are related by a Legendre transformation: 
$P= \rho\mu -u(\rho)$.
From this and the 
Bernoulli equation we see that
the pressure is equal to minus the action density:
\be
-P=   \rho \dot \phi +\frac 12 \rho (\nabla\phi)^2
+ u(\rho).
\ee
Consequently, we can write 
\be
\Pi_{ij}= \rho\partial_i\phi\partial_j\phi-
\delta_{ij}\left\{ \rho \dot \phi +\frac 12 \rho (\nabla\phi)^2
+ u(\rho)\right\}.
\label{EQ:noether}
\ee 

\section{The Acoustic Metric}

To obtain Unruh's wave equation we set
\bea
\phi&=&\phi_0 + \phi_1\nonumber\\
\rho&=&\rho_0 + \rho_1. 
\eea
Here $\phi_0$ and $\rho_0$ define the mean flow. We assume  that they obey
the equations of motion. The quantities  
$\phi_1$ and $\rho_1$
represent small amplitude perturbations. 
Expanding $S$ to quadratic order in
these perturbations gives
\be
S=S_0+ \int d^4x\left\{\rho_1 \dot \phi_1
+ \frac 12 \left(\frac {c^2}{\rho_0}\right) \rho_1^2
 +\frac 12 \rho_0 (\nabla\phi_1)^2
+ \rho_1\v\cdot\nabla \phi_1 \right\}. 
\label{EQ:first_action}
\ee
Here $\v \equiv \v_{(0)}= \nabla\phi_0$.  The speed of sound, $c$, is defined by   
\be
\frac {c^2} {\rho_0}= \left.\frac {d \mu}{d \rho}\right|_{\rho_0}, 
\ee
or more familiarly
\be
c^2= \frac {d P}{d \rho}.
\ee
The terms linear in the perturbations  vanish because of
our assumption that the zeroth-order variables obey the equation of motion.

The equation of motion for  $\rho_1$ derived from
(\ref{EQ:first_action}) is 
\be
\rho_1 = - \frac{\rho_0}{c^2}\{ \dot \phi_1+ \v\cdot\nabla
\phi_1\}.
\label{EQ:rhoeq}
\ee
Since  $\rho_1$ occurs
quadratically, we may use (\ref{EQ:rhoeq}) to eliminate it and   obtain an   
effective action for the  potential $\phi_1$ only 
\be
S_2 = \int d^4x\left\{ -\frac 12 \rho_0
(\nabla\phi_1)^2 + 
\frac {\rho_0} {2 c^2}
(\dot \phi_1+ \v\cdot\nabla \phi_1)^2\right\}.
\label{EQ:unruh_action}
\ee

The resultant equation of motion for $\phi_1$ is
\cite{unruh1,unruh2}
\be
\left(\frac{\partial}{\partial t}+\nabla\cdot\v\right)\frac
{\rho_0}{c^2}\left(\frac{\partial}{\partial t}+\v\cdot\nabla\right)\phi_1=
\nabla(\rho_0\nabla\phi_1).
\label{EQ:unruheq}
\ee
This  can be written as\footnote{I use the convention 
that Greek letters 
run over four space-time 
indices  $0,1,2,3$ with $0\equiv t$, while Roman indices refer 
to the three space
components.}
\be
\frac 1{\sqrt{-g}} \partial_\mu {\sqrt{-g}}g^{\mu\nu}\partial_\nu
\phi_1=0,
\label{EQ:scalar_eq}
\ee
where 
\be
\sqrt{-g}g^{\mu\nu} = \frac {\rho_0}{c^2}\left(\matrix{ 1, & \v^T \cr
                                                            \v,& \v\v^T - c^2{\bf
1}\cr}\right).
\label{EQ:unruh_metric_up}
\ee  

We find that $\sqrt{-g}= \sqrt{\det g_{\mu\nu}}=
{\rho_0^2}/{c}$, 
and the covariant components of the metric are 
\be
g_{\mu\nu}= \frac {\rho_0}{c}\left(\matrix{ c^2-v^2, & \v^T \cr
                                                            \v,& -{\bf
1}\cr}\right).
\label{EQ:unruh_metric_down}
\ee

The associated space-time interval can be written 
\be
ds^2= \frac {\rho_0} c \left\{c^2dt^2
-\delta_{ij}(dx^i-v^idt)(dx^j-v^jdt)\right\},
\ee
and  metrics of this form, although without the overall conformal factor
${\rho_0}/{c}$, appear in the  Arnowitt-Deser-Misner (ADM)
formalism of general relativity\cite{ADM}.  There, $c$ and $-v^i$
are referred to as  the {\it lapse function\/} and {\it shift vector\/} 
respectively.
They serve to glue successive three-dimensional time slices
together to form a four-dimensional space-time\cite{MTW}. In our
present case, provided  ${\rho_0}/{c}$ can be regarded as a constant, 
each $3$-space is ordinary flat ${\bf R^3}$ equipped with
the rectangular Cartesian metric $g^{(space)}_{ij}=\delta_{ij}$ ---
but the resultant space-time is in general curved, the
curvature depending on the degree of inhomogeneity of the mean flow
$\v$.

In the geometric acoustics limit, sound  will travel along the null
geodesics defined by $g_{\mu\nu}$. Even in the presence of spatially
varying $\rho_0$ we would expect the ray paths to depend only on
the local values of $c$ and $\v$, so it  is perhaps a bit
surprising to see the density entering the expression for
the Unruh metric. An overall conformal factor, however, does not affect  {\it null\/}
geodesics, and thus variations in $\rho_0$ do not influence the ray
tracing. 

\section{Second-order Quantities}

We are going to derive various energy and momentum
conservation laws from our wave equation. Before we do, let
us consider what sort of quantities we would want them to
contain.  

It is reasonable to define the  momentum density and the momentum flux
tensor associated with the sound field
as the second order averages
\be
{\bf j}^{\rm (phonon)}= \vev{\rho_1\v_{(1)}} + \v\vev{\rho_2},
\ee
and 
\be
\Pi_{ij}^{\rm (phonon)}= \rho_0\vev{v_{(1)i}v_{(1)j}}
+v_{i}\vev{\rho_1v_{(1)j}}+v_{j}\vev{\rho_1v_{(1)i}}
+\delta_{ij}\vev{P_2} + v_{i}v_{j}\vev{\rho_2}.
\ee
In these expressions I have taken no account of any steady
part of $\v_{(2)}$. This is not a quantity intrinsic to the
sound field and has to be found by methods outside the purely
acoustic. The other second-order quantities
$P_2$ and $\rho_2$ {\it can\/} be computed in terms of
first-order amplitudes.

For $P_2$  we combine  
\be
\Delta P= \frac{d P}{d\mu}\Delta \mu + \frac 12  \frac{d^2
P}{d\mu^2}(\Delta\mu)^2 +O((\Delta\mu)^3)
\ee
and Bernoulli's equation in the form
\be
\Delta \mu = - \dot \phi_1 - \frac 12 (\nabla \phi_1)^2 - 
\v\cdot \nabla \phi_1, 
\ee
with
\be
\frac{d P}{d\mu}= \rho,\qquad  \frac{d^2
P}{d\mu^2}=\frac{d \rho}{d\mu} =
\frac{\rho}{c^2}.
\ee
Expanding out and grouping terms of appropriate orders gives
\be
P_1= -\rho_0 (\dot\phi_1 + \v\cdot\nabla \phi_1) = c^2\rho_1,
\label{EQ:P1}
\ee
which we already knew, and
\be
P_2 =-\rho_0\frac 12 (\nabla \phi_1)^2 +
\frac 12 \frac{\rho_0}{c^2}(\dot\phi_1 + \v\cdot\nabla
\phi_1)^2.
\label{EQ:P2}
\ee
We see that $P_2=\sqrt{-g}L$ where $L$ is the  Lagrangian 
density for our sound wave equation. 
For  a plane wave
$\vev{P_2}=0$

The second order change in the density,  $\rho_2$, may be
found similarly. It  is 
\be
\rho_2= \frac 1{c^2} P_2 - \frac
1{\rho_0} \rho_1^2 \left(\frac{d\ln c}{d\ln \rho}\right)_{S}.
\label{EQ:true_rho2}
\ee
For  a plane wave the time average 
$\vev{P_2}=0$, but because $\rho_2$ contains $\rho_1^2$, the time average of
this quantity is non-zero. The resulting change of volume of
the fluid, or, if the volume is held fixed,  the resulting
pressure change, is the origin of the isotropic terms in the
radiation stress tensor discussed earlier.

\section{Conservation Laws}

Because the linear wave equation does not have access to
information about counterflows or  second-order density
changes, we will not be able to derive the real energy  and
momentum fluxes from its solution. We can still derive from
the wave equation what look superficially like conservation
laws for these quantities, however, and these laws give us
insight into the behaviour of the solutions. The conserved
quantities are of course the pseudoenergy and
pseudomomentum.       

We begin by defining  a (pseudo)-energy-momentum tensor 
\be
T^{\mu\nu}= \frac 2{\sqrt{-g}}\frac{\delta S_2}{\delta g_{\mu\nu}}.
\ee
Let us recall how such a  tensor comes to be associated with
conservation laws. Suppose  that we have an action
$S(\varphi,g_{\mu\nu})$ 
which is a functional of some field variables $\varphi(x^\mu)
$ and
the metric $g_{\mu\nu}$. If we reparameterize spacetime
so that the point that had coordinates  $x^\mu$ is now
denoted by 
$x^\mu+\epsilon^\mu$, then we have  $\varphi\to \varphi+ \delta
\varphi$ and $g_{\mu\nu}\to  g_{\mu\nu}+ \delta g_{\mu\nu}$,
where
\bea 
\delta\varphi &=& \epsilon^\mu \partial_\mu
\varphi\nonumber\\
\delta g_{\mu\nu} &=& D_\mu \epsilon_\nu +  D_\nu
\epsilon_\mu.
\eea
Here  $D_\mu$ is the covariant
derivative containing the Riemann connection compatible with
the  metric. The variation in the metric comes from the computation
\bea
ds^2&=& g_{\mu\nu}(x)dx^\mu dx^\nu\nonumber\\
    &\to& g_{\mu\nu}(x^\alpha+\epsilon^\alpha)d(x^\mu+\epsilon^\mu)
    d(x^\nu+\epsilon^\nu)\nonumber\\
    &=& (\epsilon^\alpha\partial_\alpha g_{\mu\nu}+
    g_{\alpha\nu}\partial_\mu \epsilon^\alpha+
     g_{\mu\alpha}\partial_\nu
     \epsilon^\alpha)dx^\mu dx^\nu\nonumber\\
    &=& (D_\mu \epsilon_\nu +  D_\nu \epsilon_\mu)dx^\mu
    dx^\nu.
\eea    
The assembly of the terms into  covariant derivatives
in the last line is most easily established by using geodesic
coordinates and the fact that $\delta g_{\alpha\nu}$ is a
tensor. The combination  $D_\mu \epsilon_\nu +  D_\nu
\epsilon_\mu$ is the Lie derivative, ${\cal L}_\epsilon
g_{\mu\nu}$, of the metric with
respect to the vector field $\epsilon^\mu$.

Since  a mere re-coordinatization  does not
change the numerical value of the action, we must have
\be
0=\delta S = \int d^4x \sqrt{-g}\left\{(D_\mu \epsilon_\nu +  D_\nu
\epsilon_\mu)\frac {1}{\sqrt{-g}}\frac{\delta S}{\delta
g_{\mu\nu}}  
+ (\epsilon^\mu\partial_\mu \varphi) \frac{1} {\sqrt{-g}}\frac{\delta
S}{\delta\varphi}\right\}.
\ee  
Now the equations of motion state that $S$ is unchanged by
any variation in $\varphi$, including, {\it a fortiori\/}
the change $\delta\varphi = \epsilon^\mu \partial_\mu
\varphi$. 
Thus 
\be 
0=\int d^4x\,\sqrt{-g}\, (D_\mu \epsilon_\nu)\frac {2}{\sqrt{-g}}\frac{\delta S}{\delta
g_{\mu\nu}} = \int d^4x\,\sqrt{-g}\, \epsilon_\nu
D_\mu\left(\frac
{2}{\sqrt{-g}}\frac{\delta S}{\delta
g_{\mu\nu}}\right),
\label{EQ:depsilon}
\ee
where, in the last equality, we have integrated by parts by using the
derivation property of the covariant derivative and the
expression
\be D_\mu J^\mu=\frac  1{\sqrt{-g}}
\partial_\mu (\sqrt{-g}J^\mu)
\label{EQ:divergence}
\ee 
for
the divergence of a vector.  
Since  (\ref{EQ:depsilon}) is true for arbitrary
$\epsilon^\mu(x)$, we deduce that
\be
D_\mu\left(\frac
{2}{\sqrt{-g}}\frac{\delta S}{\delta
g_{\mu\nu}}\right)=D_\mu T^{\mu\nu}=0.
\label{EQ:covcons}
\ee

Although (\ref {EQ:covcons}) has the appearance of a
conservation law, and has useful applications in itself, 
we have not yet exploited  any symmetries of the
system --- and it is symmetries that lead to conserved
quantities. To derive a genuine local conservation law  
we need to assume that the metric admits a Killing
vector, $\eta^\mu$. This means that the particular reparameterization
$\epsilon^\mu=\eta^\mu$ 
is actually an isometry of the manifold and so leaves the
metric invariant    
 \be
{\cal L}_\eta g_{\mu\nu} = D_\mu\eta_\nu+ D_\nu\eta_\mu=0.
\label{EQ:killing}
\ee

Combining (\ref{EQ:killing}) 
with (\ref{EQ:covcons}) and using the symmetry of
$T^{\mu\nu}$  we find that
\be
D_\mu (T^{\mu\nu}\eta_\nu)=0.
\ee
Using (\ref{EQ:divergence}), this can be written 
\be
\partial_\mu (\sqrt{-g}\, T^{\mu\nu}\eta_\nu)=0.
\ee
Thus   the 4-vector density $Q^\mu= \sqrt{-g}\,
T^{\mu\nu}\eta_\nu$ is conventionally conserved, and 
\be
Q=\int d^3x Q^0
\ee
is independent of the time slice on which it is evaluated.

Since (\ref{EQ:unruh_action}) can be written as the usual
action for a scalar field  
\be
S_2=\int d^4x \frac 12 \sqrt{-g} g^{\mu\nu}\partial_\mu
\phi_1\partial_\nu\phi_1,
\ee
we have 
\be
T^{\mu\nu}= \partial^\mu \phi_1\partial^\nu \phi_1-
g^{\mu\nu}\left(\frac 12 g^{\alpha\beta}
\partial_\alpha\phi_1\partial_\beta\phi_1\right).
\label{EQ:Tmunu}
\ee
The derivatives with raised indices in (\ref{EQ:Tmunu}) are
defined by
\be
\partial^0 \phi_1 = g^{0\mu}\partial_\mu\phi_1= \frac
1{\rho_0c} (\dot\phi_1 + \v\cdot\nabla \phi_1),
\ee
and
\be
\partial^i \phi_1=g^{i\mu}\partial_\mu\phi_1=   
\frac
1{\rho_0c} \left( v_{i}(\dot\phi_1 + \v\cdot\nabla \phi_1)
   -c^2\partial_i \phi_1\right).
\ee

Thus 
\bea
T^{00}&=& \frac{1}{\rho_0^3}\left(\rho_0\frac 12 (\nabla \phi_1)^2 +
\frac 12 
\frac{\rho_0}{c^2}(\dot\phi_1 + \v\cdot\nabla
\phi_1)^2\right)\nonumber\\ 
&=& \frac{c^2}{\rho_0^3} \left(\frac {{\cal E}_{r}}{c^2}\right) \nonumber\\
&=&\frac{c^2}{\rho_0^3}\tilde\rho_2.
\label{EQ:TOO}
\eea 
The last two equalities serve as a definition of  ${\cal E}_{r}$ and
$\tilde\rho_2$.  The quantity  ${\cal E}_{r}$ is often   described as the
acoustic energy density relative to the frame moving with the local
fluid velocity\cite{lighthill}. It is, of course,  more correctly a 
pseudo-energy density.

We can express the other components  of (\ref{EQ:Tmunu}) in terms of
physical quantities.  We  find that
\bea
T^{i0}=T^{0i}&=& \frac{c^2}{\rho_0^3}\left(\frac 1{c^2} (P_1v_{(1)i} +
v_{i}{\cal E}_{r})\right)\cr
&=&\frac{c^2}{\rho_0^3}\left( \rho_1v_{(1)i}+
v_{i}
\tilde\rho_2\right).
\eea
The first  line in this expression shows that, up to an overall
factor, $T^{i0}$ is an  energy flux --- the first term being the rate
of working  by a fluid element on its neighbour, and the second the
advected energy.  The second  line is written so as to suggest the
usual relativistic identification  of (energy-flux)$/c^2$ with  the
density of momentum.  This  interpretation, however, requires that
$\tilde\rho_2$ be  the  second-order correction to the density,
which, sadly, it is not.

Similarly
\be
T^{ij}=\frac{c^2}{\rho_0^3}\left(\rho_0v_{(1)i}v_{(1)j}
+v_{i}\rho_1v_{(1)j}+v_{j}\rho_1v_{(1)i}
+\delta_{ij}P_2 + v_{i}v_{j}\tilde\rho_2\right).
\ee
We again see that if we were only able to identify $\tilde\rho_2$ with $\rho_2$
then $T^{ij}$ has  precisely  the  form   we expect for the
second-order momentum flux tensor. Although it comes close, 
the inability of the
pseudomomentum flux to  exactly  capture the true-momentum flux is
inevitable as we know that computing the true stresses in the medium 
requires more information about the equation of state than is
available to the linearized wave equation.

We can also write the mixed co- and contra-variant components of the energy 
momentum tensor  $T^{\mu}_{\phantom {\mu}\nu}=
T^{\mu\lambda}g_{\lambda\nu}$
in terms of physical quantities. This mixed tensor  turns out to be  
more useful than the doubly contravariant tensor.  Because 
we no longer  enforce a symmetry  between the indices $\mu$ and $\nu$, 
the quantity ${\cal E}_{r}$ is no longer
required to perform double duty as both an energy and a density. We find   
\bea
\sqrt{-g}\,T^{0}_{\phantom {0}0} &=& \left( {\cal E}_{r}+
\rho_1\v_{(1)}\cdot \v\right)\cr
\sqrt{-g}\,T^{i}_{\phantom {i}0}&=& 
\left( \frac {P_1}{\rho_0} + 
\v\cdot \v_{(1)}\right)(\rho_0v_{(1)i}+\rho_1v_{(0)i}),
\eea 
and 
\bea
\sqrt{-g}\,T^{0}_{\phantom {0}i}&=& -\rho_1\v_{(1)i}\cr
\sqrt{-g}\,T^{i}_{\phantom {i}j}&=& -\left(\rho_0v_{(1)i}v_{(1)j}
+ v_i\rho_1v_{(1)j} + \delta_{ij}P_2\right). 
\eea
We see that $\tilde \rho_2$ does not appear here, and all these
terms may be identified with physical quantities which are  reliably computed 
from solutions of the linearized wave equation.

Now we turn to the local conservation laws.
In what follows I will consider only a steady  background flow, and further one 
for which $\rho_0$, $c$, and hence 
$\sqrt{-g}=\rho_0^2/c$ can be treated as constant.
To increase the readability of some  expressions I will also
choose units so that $\rho_0$ and $c$ become unity and no longer
appear as overall factors in the metric or the four-dimensional
energy-momentum tensors. I will, however, reintroduce them when they
are required for  dimensional correctness in expressions such as
$\rho_0\v_{(1)}$ or ${\cal E}_{r}/c^2$.

From the acoustic metric  we find
\bea
\Gamma^0_{00}&=& \frac 12 (\v \cdot \nabla) |v|^2 \cr
\Gamma^0_{i0}&=& -\frac 12 \partial_i\, |v|^2 +
\frac 12 v_j(\partial_i v_j-\partial_j v_i)\cr
\Gamma^i_{00}&=& \frac 12 v_{i} (\v \cdot \nabla) |v|^2  - 
\frac 12 \partial_i\, |v|^2  \cr
\Gamma^0_{ij}&=&  \frac 12 (\partial_i v_{j} + \partial_j v_{i}) \cr
\Gamma^i_{j0}&=&  -\frac 12 v_{i}\partial_j\, |v|^2  
+\frac 12 (\partial_j v_k - \partial_k v_j)(v_kv_i-c^2\delta_{ik}) \cr
\Gamma^i_{jk}&=&  \frac 12  v_{i}(\partial_j v_{k} + \partial_k
v_{j}).
\eea

We now  evaluate 
\bea
D_\mu T^{\mu 0} &=&  \partial_\mu  T^{\mu
0} + \Gamma^\mu_{\mu\gamma} T^{\gamma 0}+\Gamma^0_{\mu\nu} T^{\mu\nu}\cr
   &=& \partial_\mu T^{\mu
0} + \Gamma^0_{\mu\nu} T^{\mu\nu}.
\eea
After a little algebra we find 
\be
 \Gamma^0_{\mu\nu} T^{\mu\nu} = \frac 12 (\partial_i v_j +
\partial_j v_i)(\rho_0 v_{(1)i} v_{(1)j}+ \delta_{ij}P_2). 
\ee
Note the  non-appearance of  $\rho_1$
and $\tilde\rho_2$ in the final expression --- even though 
both quantities appear in
$ T^{\mu\nu}$.

The conservation law therefore becomes 
\be
\partial_t {\cal E}_{r} +\partial_i (P_1v_{(1)i}+v_i{\cal E}_{r}) + \frac 12 \Sigma_{ij} 
(\partial_i v_j +
\partial_j v_i)=0,
\ee
where
\be
\Sigma_{ij}= \rho_0 v_{(1)i} v_{(1)j}+ \delta_{ij}P_2.
\ee
This is  an example of the general form of  energy law
derived by Longuet-Higgins and Stuart, originally in the
context of ocean
waves\cite{longuet-higgins1}.   The
relative energy density, ${\cal E}_{r}\equiv T^{00}$, is {\it not\/}
conserved. Instead, an observer moving with the fluid  sees
the waves acquiring energy from the mean flow at a rate
given by the product of a radiation stress, $\Sigma_{ij}$,
with the mean-flow rate-of-strain. 
This equation is sometimes cited\cite{longuet-higgins2} as
an explanation for the monstrous ship-destroying  waves  
that may be encountered
off the eastern  coast of South Africa. Here long
wavelength swell from distant Antarctic storms runs into the
swift southbound  Agulhas current and is greatly amplified
by the opposing flow.  

We  now examine  the energy conservation law coming from the zeroth 
component of the mixed  
energy-momentum tensor.  After a little work we find that 
the  connection  contribution vanishes identically and  
the energy conservation law
is therefore
\be
\partial_t\left( {\cal E}_{r}+
\rho_1\v_{(1)}\cdot \v\right)+ \partial_i\left( (\frac {P_1}{\rho_0} + 
\v\cdot \v_{(1)})(\rho_0v_{(1)i}+\rho_1v_{(0)i})\right)=0.
\label{EQ:blokhintsev}
\ee
We see that the combination $ {\cal E}_{r}+ \rho_1\v_{(1)}\cdot \v$
{\it does \/} correspond to a conserved energy --- as we
should have anticipated since a steady flow provides us with
a Killing vector ${\bf e}_0=\partial_t$.  This 
conservation law was originally derived  by
Blokhintsev\cite{blokhintsev}  for slowly varying flows, and more
generally by Cantrell and Hart\cite{cantrell} in their study of the
acoustic stability of rocket engines. 

Finally the   covariant conservation equation  $D_\mu T^{\mu}_{\phantom{\mu}j}=0$ 
reads 
\be
\partial_t \rho_1v_{(1)j} + \partial_i \left(\rho_0v_{(1)i}v_{(1)j}
+ v_i\rho_1v_{(1)j} + \delta_{ij}P_2\right)
+ \rho_1v_{(1)i}  \partial_j v_i=0.
\label{EQ:momnoncon}
\ee
Here  connection terms  have provided a source term for the momentum density.
Thus, in an inhomogeneous flow field,  momentum is exchanged between the
waves and the mean flow.

\section{Phonons and Conservation of Wave Action}

The conservation laws we have derived in the previous
section may all be interpreted in terms of the
semiclassical motion of phonons. As noted by McIntyre\cite{mcintyre81}, 
the existence of such an interpretation  is a sure sign 
that the conservation laws in question are those
of pseudomomentum and pseudoenergy. 

To make contact with the semiclassical picture we observe
that  when the mean flow varies  slowly on the scale of a
wavelength, the sound  field can  locally be approximated
by  a plane wave  
\be
\phi(x,t)= A \cos({\bf k}\cdot {\bf x} -\omega t).
\ee
The frequency $\omega$ and the wave-vector $\bf k$ are  related by the 
Doppler-shifted dispersion relation 
$
\omega= \omega_r + {\bf k}\cdot{\v},
$
where   $\omega_r=c|k| $, is the frequency relative to  
a frame moving with the
fluid.  A packet of such waves 
is governed by Hamilton's ray equations 
\be
\frac {dx^i}{dt}= \frac {\partial \omega}{\partial
k^i},\qquad
\frac {dk^i}{dt}= -\frac {\partial \omega}{\partial x^i}.
\ee
In other words the packet moves at the group velocity 
\be
{\bf V}_g =\dot {\bf x}=  c \frac{{\bf k}}{|k|} +\v.
\ee
and the change in  ${\bf k}$ 
is given by 
\be
\frac{d k_j}{dt} =- k_i \frac{\partial v_i}{\partial
x^j}.
\label{EQ:kdot}
\ee
In this equation  the  time derivative is taken along the ray:
\be
 \frac{d }{dt} = \frac{\partial}{\partial t} + {\bf V}_g\cdot \nabla.
\ee   
This  evolution  can also be derived from the equation for  null geodesics of the
acoustic metric\cite{stone00b}.

The evolution of the amplitude $A$ is linked with that of the relative energy
density, ${\cal E}_{r}$, through
\be
\vev{{\cal E}_{r}}= \frac 12 A^2 \rho_0
\frac{\omega_r^2}{ c^2}.
\ee
For a  homogeneous stationary fluid we would expect our    
macroscopic plane wave to correspond to  a  quantum coherent state whose energy 
is, in terms of  the (quantum) average phonon density $\bar N$,
\be
E_{tot}= (\hbox{\rm Volume}) \vev{ {\cal E}_{r}} =(\hbox{\rm Volume}) \bar N\hbar \omega_r.    
\ee
Since it is a density of ``particles'', 
$\bar N$ should remain the same when viewed from any frame. Consequently,   the relation
\be
\bar N\hbar= \frac{\vev{{\cal E}_{r}}}{\omega_r}
\ee
should hold true generally. In classical fluid mechanics the quantity   
$\vev{{\cal E}_{r}}/\omega_r$ is called 
the {\it wave action\/}\cite{garrett,lighthill,andrews79b}.

The time averages of  other components of the energy-momentum 
tensor may be also expressed in terms of $\bar N$.
For the mixed tensor we find
\bea
\vev{\sqrt{-g}\,T^0_{\phantom{0}0}}&=&\vev{{\cal E}_{r} + \v\cdot \rho_1 \v_{(1)}}  = \bar N\hbar
\omega\cr
\vev{\sqrt{-g}\,T^i_{\phantom{i}0}} &=& \vev{( \frac {P_1}{\rho_0} +\v\cdot
\v_{(1)})( \rho_0v_{(1)i}+\rho_1v_i)} =\bar N\hbar \omega
(V_g)_i\cr
\vev{-\sqrt{-g}\,T^0_{\phantom{0}i}}&=&\vev{\rho_1 v_{(1)i} }=  \bar N\hbar 
k_i\cr
  \vev{-\sqrt{-g}\,T^i_{\phantom{i}j}} &=&  \vev{ \rho_0 v_{(1)i}v_{(1)j} 
+ v_i  \rho_1 v_{(1)j} + \delta_{ij} P_2} =  \bar N\hbar k_j  (V_g)_i.
\eea
In the  last equality we have used that  $\vev{P_2}=0$ 
for a plane progressive wave.

If we insert these expressions for the time averages into the
Blokhintsev energy  conservation law (\ref{EQ:blokhintsev}), we find that
\be
\frac{\partial \bar N\hbar \omega}{\partial t} + \nabla\cdot 
(\bar N\hbar \omega {\bf V}_g)=0.
\ee
We can write this as
\be
\bar N\hbar\left(\frac{\partial  \omega}{\partial t}+ {\bf
V}_g\cdot \nabla
\omega\right)
+ \hbar\omega\left( \frac{\partial \bar N }{\partial t} + \nabla\cdot 
(\bar N {\bf V}_g)\right)=0.
\ee  
The first term is proportional to  ${d\omega}/{dt}$ taken
along the rays and   vanishes for a steady mean flow as a
consequence of the Hamiltonian nature of the ray tracing
equations. The second term must therefore also vanish. This
vanishing represents the conservation of phonons, or, in
classical language, the conservation of wave-action.
Conservation of wave action is an analogue of the  
adiabatic invariance of $E/\omega$ in the time dependent
harmonic oscillator.

In a similar manner, the time average of (\ref{EQ:momnoncon}) may be written
\bea
0 &=& \frac{\partial \bar N k_j}{\partial t} + \nabla\cdot 
(\bar N k_j {\bf V}_g) + \bar N k_i \frac{\partial v_i}{\partial
x^j}\cr
 &=& \bar N\left(\frac{\partial k_j}{\partial t}+ {\bf V}_g\cdot \nabla
k_j+ k_i\frac{\partial v_i}{\partial
x^j} \right)
+  k_j\left( \frac{\partial \bar N }{\partial t} + \nabla\cdot 
(\bar N {\bf V}_g)\right).
\eea
We see that the momentum law becomes equivalent to  phonon-number conservation 
combined with the ray tracing equation
(\ref{EQ:kdot}).

\section{Summary}

When dealing with waves in a medium it is essential to
distinguish between  the conceptually distinct quantities
of momentum and pseudomomentum. Under suitable
circumstances either may be used for computing forces ---
but there is no general rule for determining those
circumstances. In the case of non-dispersive sound waves,
momentum and pseudomomentum often travel together, and are
therefore easily confused --- but the conservation laws
derived by manipulating the wave equation are those of
pseudomomentum and pseudoenergy. The GR analogy provided by
the acoustic metric provides a convenient  and structured
route to deriving these laws when the background fluid is
moving.

\section{Acknowledgements}

This work was supported by  NSF grant number DMR-98-17941. 
I would like to thank the staff and members of NORDITA in Copenhagen,
Denmark,  for 
their hospitality while this paper was being written.

\eject
\end{document}